\documentclass[preprint,showpacs,preprintnumbers,amsmath,amssymb]{revtex4}
\usepackage{dcolumn}
\usepackage{bm}
\usepackage{graphicx}
\begin{document}

\title{High-Frequency Nanofluidics: An Experimental Study using Nanomechanical Resonators} \date{\today}

\author{D. M. Karabacak}
\affiliation{Department of Aerospace and Mechanical Engineering, Boston University, Boston, Massachusetts, 02215}
\author{V. Yakhot}
\affiliation{Department of Aerospace and Mechanical Engineering, Boston University, Boston, Massachusetts, 02215}
\author{K. L. Ekinci\footnote{Author to whom correspondence should be addressed. Electronic mail: ekinci@bu.edu}}
\affiliation{Department of Aerospace and Mechanical Engineering, Boston University, Boston, Massachusetts, 02215}

\begin{abstract}

Here we apply nanomechanical resonators to the study of oscillatory fluid dynamics.  A high-resonance-frequency
nanomechanical resonator generates a rapidly oscillating flow in a surrounding gaseous environment; the nature of the
flow is studied through the flow-resonator interaction. Over the broad frequency and pressure range explored, we
observe signs of a transition from Newtonian to non-Newtonian flow at $\omega\tau\approx 1$, where $\tau$ is a properly
defined fluid relaxation time. The obtained experimental data appears to be in close quantitative agreement with a
theory that predicts purely elastic fluid response as $\omega\tau\rightarrow \infty$.

\end{abstract}
\maketitle

The Navier-Stokes equations based upon the Newtonian approximation have been remarkably successful over the centuries
in formulating solutions for relevant flow problems both in bulk and near solid walls \cite{landaulifshitz}. The
Newtonian approximation breaks down, however, when the particulate nature of the fluid becomes significant to the flow.
The Knudsen number, $Kn=\lambda /L$, is one parameter, which is commonly used to settle whether the Newtonian
approximation can be applied to a medium or not. Here, one compares the mean free path $\lambda$ in the medium to an
{\it ill-defined} characteristic length $L$. A second defining parameter, especially for oscillatory flow, is the
Weissenberg number, $Wi=\tau/T$, which compares the characteristic time scale $T$ of the flow with the relaxation time
$\tau$ in the medium. As $\tau/T=\omega\tau$ is varied --- for instance, by varying the flow frequency $\omega$ or the
relaxation time $\tau$ --- the nature of the flow changes drastically.

Recent developments in nanometer scale engineering have created a vibrant subfield of fluid dynamics called
nanofluidics \cite{nanoflows}. Most nanofluidics work is concerned with flow in nanoscale channels and remains strictly
in the Newtonian regime. In contrast, emerging nanometer scale mechanical resonators \cite{SciAm,ekinci-nems}, with
frequencies already extended into the microwaves \cite{roukes-ghz,zettl-ghz}, offer an uncharted parameter space for
studying nanofluidics. For a high-frequency nanomechanical resonator with resonance frequency $\omega/2\pi$, one can
tune $\omega\tau$ over a wide range --- in fact, possibly reaching the limits of the Newtonian approximation in a given
liquid or gas. This not only allows experimental probing of a flow regime that was inaccessible by past experiments
\cite{viscometer-measurement,qcm-rodahl}, but also presents the unique prospect of designing nanodevices for key
technological applications.

To complement the recent theoretical interest in high frequency nanofluidics
\cite{yakhot-stokes,bhiladvala,nanoflows,paul-stochastic}, we experimentally studied the interaction of high-frequency
nanomechanical resonators with a gaseous environment. The gaseous environment presents an ideal fluid for these
studies, where one can effectively tune $\tau$ by changing the pressure $p$. On the other hand, varying the resonator
dimensions changes the mechanical resonance frequency $\omega/2\pi$. When combined, the two experimental parameters
allow $\omega\tau$ to be varied over several orders of magnitude effectively.

\begin{figure}
\includegraphics[trim=5mm 6mm 5mm 7mm,clip=true,scale=.63]{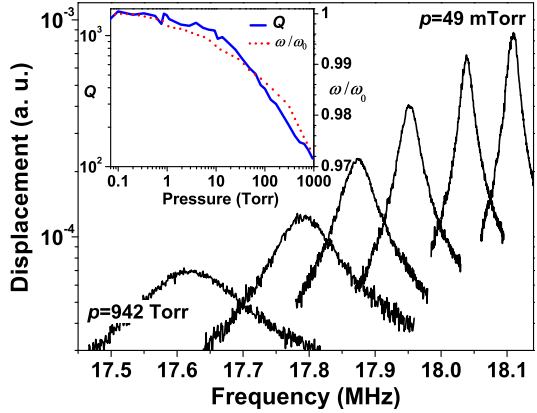}
\caption{Resonance of a silicon doubly-clamped beam of width $w=500$ nm, thickness $h=280$ nm and length $l=11.2$
$\mu$m at various N$_2$ pressures in the chamber: $p=$ 0.049, 5.4, 32, 100, 302 and 942 Torr.  The inset shows the
extracted quality factor $Q$ and normalized resonance frequency $\omega/\omega_0$ of the same device as a function of
pressure. Here, $Q_0\approx 1530$.} \label{signal}
\end{figure}

In order to cover a broad frequency range, we fabricated silicon doubly-clamped beam resonators with varying dimensions
$w\times h\times l$ displayed in Table~{\ref{tab:devices}} using standard techniques \cite{kouh-apl}. To further extend
the frequency range, we employed fundamental and first harmonic modes of two commercial silicon AFM cantilevers
(Table~{\ref{tab:devices}}). For the measurements, we used a pressure-controlled optical characterization chamber
connected to a high purity N$_2$ source. We actuated the out-of-plane modes of the resonators  electrostatically
and measured the displacements optically \cite{kouh-apl}. All measurements were performed under linear drive; moreover,
the results remained independent of the rms displacement amplitudes of 1-10 nm as confirmed by Michelson
interferometry.
\begin{table}
\caption{\label{tab:devices} Device parameters, transition pressure $p(Wi=1)$ and the approximate lower pressure limit
$p_{min}$ for accurate measurements for the devices used in the study. 1$^{\textrm{st}}$ harmonic mode was also
employed for some AFM cantilevers.}
\begin{ruledtabular}
\begin{tabular}{ccccc}
$\left(w \times h \times l\right)$&$\omega_0/2\pi$& $Q_0$ & $p(Wi=1)$& $p_{min}$\\
 ($\mu$m) & (MHz) & & (Torr) & (Torr) \\
\hline
$53\times2\times460$ (1$^{\textrm{st}}$ Harmonic) & $0.078$ & $8321$ & $1.0$ &$0.05$\\
$36\times3.6\times125$ (Fundamental) & $0.31$ & $8861$ & $3.0$ &$0.05$\\
$36\times3.6\times125$ (1$^{\textrm{st}}$ Harmonic) & $1.97$ & $3522$ & $17.5$ &$0.06$\\
$0.50\times0.28\times17.1$ & $10.4$ & $1840$ & $110$ &$1.9$\\
$0.50\times0.28\times11.2$ & $18.1$ & $1530$ & $200$ &$2.9$\\
$0.93\times0.22\times9.9$ & $22.8$ & $1335$ & $176$ &$0.8$\\
$0.76\times0.22\times9.9$ & $22.9$ & $1200$ & $216$ &$1.2$\\
$0.23\times0.20\times9.6$ & $24.2$ & $415$ & $280$ &$2.77$\\
$0.50\times0.28\times9.1$ & $27.1$ & $909$ & $290$ &$2.6$\\
$0.32\times0.20\times7.7$ & $33.2$ & $780$ & $320$ &$15.8$\\
$0.50\times0.28\times5.9$ & $45.7$ & $1066$ & $310$ &$2.3$\\
$0.25\times0.20\times5.6$ & $53.2$ & $571$ & $400$ &$1.2$\\
$0.73\times0.23\times5.6$ & $58.6$ & $525$ & $490$ &$19.0$\\
$0.24\times0.20\times3.6$ & $102.5$ & $495$ & $-$ &$11.9$\\
\end{tabular}
\end{ruledtabular}
\end{table}
Figure~{\ref{signal}} depicts the typical resonant response of a nanomechanical resonator as the background N$_2$
pressure in the chamber is increased. The frequency shift is due to the mass loading from the boundary layer
\footnote{Virtual mass effect due to the potential flow around the structure was determined to be negligible.}, while
the broadening results from the energy dissipation in the fluid. The analysis can be simplified by using a
one-dimensional damped harmonic oscillator approximation \cite{nanomechanics}, $\ddot{x}+\gamma \dot{x}+\omega^2x=f/m$,
where $f/m$ represents the force per unit effective mass of the resonator. The quality factor $Q$, which is a
comparison of the stored energy to the dissipated energy per cycle, is related to $\gamma$ as $\gamma \approx \omega/Q
$. Here, we extracted both the resonance frequency $\omega/2\pi$ and $Q$ using nonlinear least squares fits to the
Lorentzian response of the resonator. In addition, for low-$Q$ (high pressure), we verified the Lorentzian fit results
through fits to the real and imaginary components of the complex transmission \cite{fitting-methods}. Typical changes
in $\omega $ and $Q$ of a nanomechanical resonator during a pressure sweep are shown in the inset of
Fig.~{\ref{signal}}. Both $\omega $ and $Q$ approach their respective {\it intrinsic} values, $\omega_0$ and $Q_0$, at
low pressure.

Before presenting further results, we must clarify the nature of the fluidic energy dissipation. The motion of the
fluid with respect to the solid boundary creates a complex, position-dependent shear stress on the resonator surface.
The inertial and dissipative components of the net shear force are proportional to the displacement and the velocity,
respectively.  For a single device, as the pressure is changed, the fluidic dissipation can be quantified by either the
fluidic quality factor $Q_f$ given by $Q_f^{-1}=Q^{-1}-Q_0^{-1}$ or the fluidic dissipation $\gamma_f=\omega /Q_f$. To
compare different devices with varying sizes and geometries, one needs to further realize that the fluidic dissipation
is proportional to the effective surface area $S_{eff}$, while the stored energy in the resonator is proportional to
the effective mass $m$ \footnote{The typical mass loading in these experiments is small. Thus, for all practical
purposes, $\omega \approx \omega_0$.}. With this naive assumption, we define a normalized fluidic dissipation,
$\gamma_n=\gamma_f m/S_{eff}$. The lower pressure limit $p_{min}$ for accurate $Q_f$ measurement is set by $Q_0$: as
one approaches $p_{min}$, the intrinsic losses in the resonator dominate the measurement. The upper limit is 1000 Torr.
Table ~{\ref{tab:devices}} displays  $p_{min}$  for each device along with $Q_0$.

\begin{figure*}
\includegraphics[trim=14mm 17mm 10mm 21mm,clip=true,scale=.667]{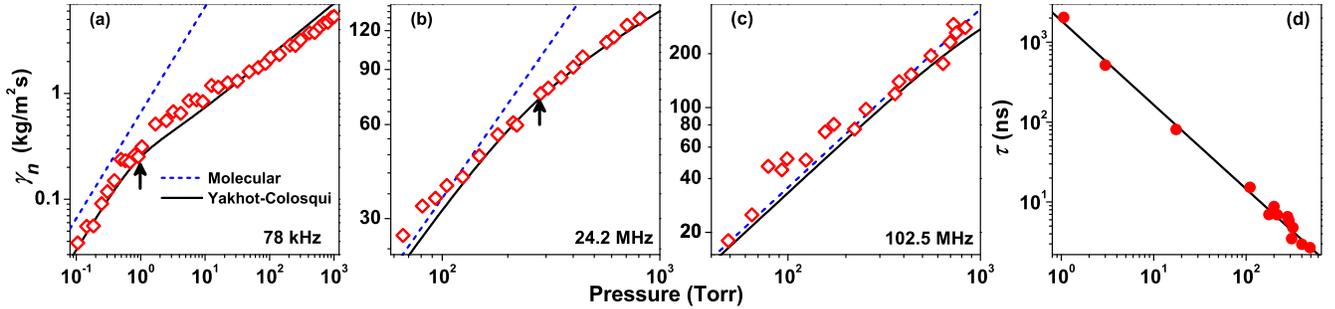}
\caption{Normalized fluidic dissipation $\gamma_n$ as a function of pressure for (a) a cantilever with dimensions
$\left(w\times h\times l\right)$ $53\times2\times460$ $\mu$m, and nanomechanical doubly-clamped beams with dimensions
of (b) $230$ nm$\times200$ nm$\times9.6$ $\mu$m  (c) $240$ nm$\times 200$ nm$\times3.6$ $\mu$m. Resonance frequencies
are as indicated and the approximate turning points are marked with arrows. The lines are fits to molecular collision
model \cite{bhiladvala} and Eq.~{\ref{gamma_eq}} using $\tau\approx1850/p$ \cite{yakhot-stokes}. The molecular
collision and the Yakhot-Colosqui predictions were multiplied by $0.9$ and $2.8$, respectively, for {\it all}
resonators. (d) Relaxation time $\tau$ as a function of pressure. The points were extracted from fluidic dissipation
data sets, such as those shown in (a) and (b), of 13 resonators. The solid line is a least-mean-squares fit and
indicates that $\tau\approx1850/p$.}\label{gamma-p}
\end{figure*}

\begin{figure}[h]
\includegraphics[trim=5mm 7mm 5mm 5mm,clip=true,scale=.6]{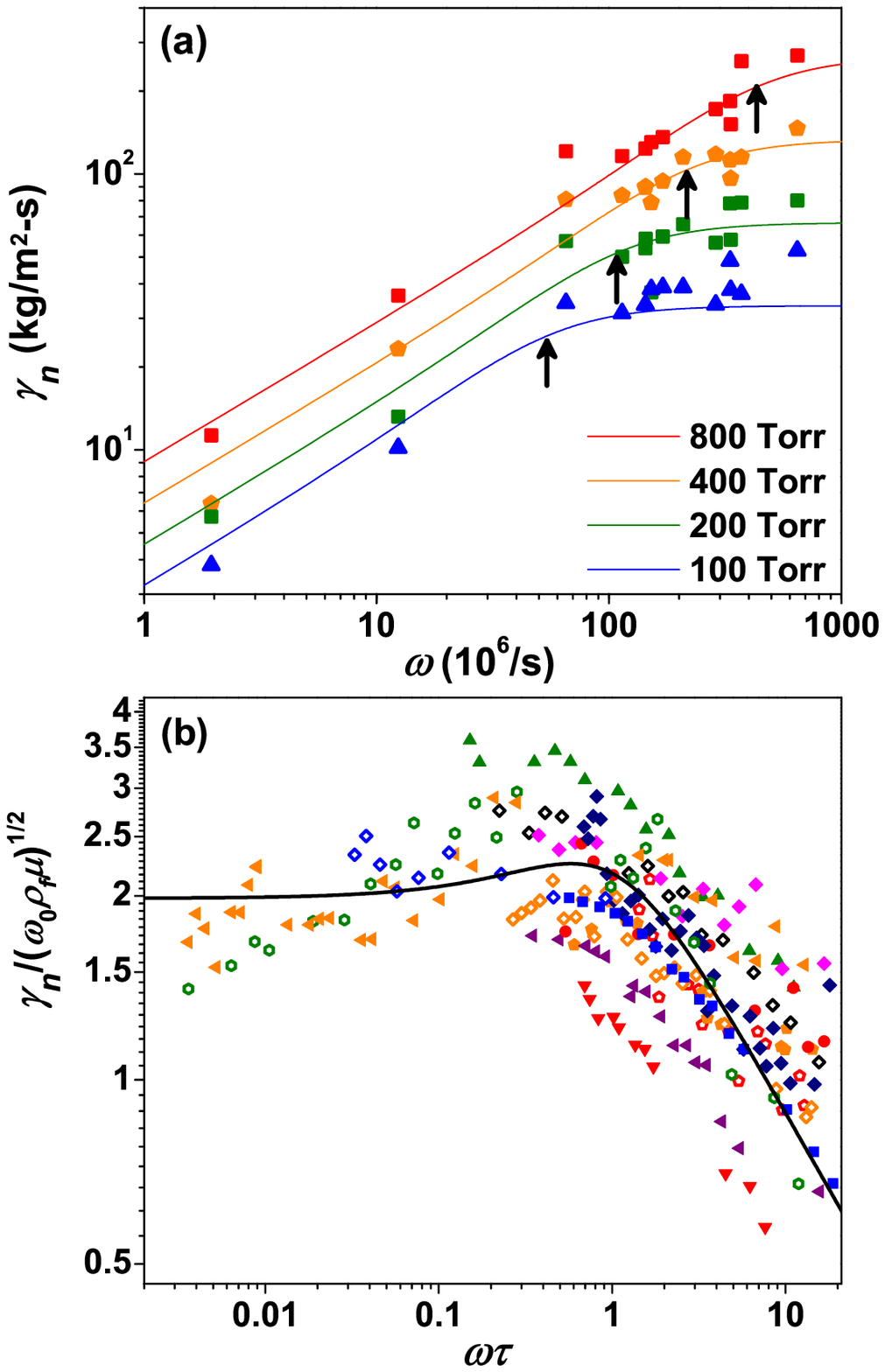}
\caption{(a) Normalized fluidic dissipation $\gamma_n$ as a function of the resonator frequency $\omega_0$ for several
resonators at four different pressures. From top to bottom, $\tau\approx$2.3, 4.6, 9.2, 18.5 ns. The lines are fits
calculated using Eq.~{\ref{gamma_eq}}. $Wi \approx 1$ points are marked with an arrow for each pressure. (b)The scaling
of  $\gamma_n/\sqrt{\omega\mu\rho_f}$ for all resonators with $\omega \tau$. Each symbol corresponds to an individual
resonator. {\it All} the predictions were multiplied by the same fitting factor of $2.8$ (also see
Fig.~{\ref{gamma-p}}).} \label{gamma-omega}
\end{figure}

The normalized fluidic dissipation $\gamma_n$ observed in three different resonators are presented in
Fig.~{\ref{gamma-p}}(a),(b) and (c) as a function of gas pressure. A change in the slope is noticeable for the data in
(a), and (b) at approximate pressures of $1$ and $300$ Torr, respectively. The turn points marked in the plots
correspond to $\omega\tau\approx1$, and is discussed in detail below. For the highest frequency beam at 102.5 MHz shown
in (c), the turn point falls outside the available pressure range, i.e., 1000 Torr. Molecular flow model
\cite{bhiladvala}, which takes into account specular collisions, fits our data only at the ideal gas limit at low
pressure. Note that a multiplicative constant of $0.9$ was used in all three to improve the fits. Viscous effects
\cite{sader,blom} and squeeze-film effects \cite{squeezefilm-damping, squeezefilm}, commonly observed in MEMS, did not
introduce significant damping for the small high-frequency devices up to atmospheric pressure.

The solid lines in Fig.~{\ref{gamma-p}} are fits to a theory by Yakhot and Colosqui \cite{yakhot-stokes} developed from
the Boltzmann Equation in the relaxation time approximation. After geometric normalization and imposing the no-slip
boundary condition, this theory culminates in the expression
\begin{eqnarray}\label{gamma_eq}
&& \gamma_n  \approx \frac{1}{\left( {1 + \omega ^2 \tau ^2 } \right)^{3/4} } \sqrt{\frac{\omega\mu\rho_f}{2}} \left[
\left( {1 + \omega \tau } \right)\right. \nonumber\\ && \left. \cos \left( \frac{\tan ^{ - 1} {\omega \tau }}{2}\right)
- {\left( {1 - \omega \tau } \right)\sin \left( {\frac{\tan ^{ - 1} {\omega \tau }}{2}} \right)} \right].
\end{eqnarray} In Eq.~{\ref{gamma_eq}}, $\gamma_n$ is expressed in terms of the viscosity $\mu$, the density $\rho_f $, the
effective relaxation time $\tau$ of the fluid, and the frequency $\omega$ of the resonator. The only unavailable
parameter is the relaxation time $\tau$.  In order to obtain the fits, we assumed that $\tau$ satisfied the empirical
form, $\tau \propto1/p$ \cite{qcm-rodahl, krim-slip}. The key prediction of the Yakhot-Colosqui \cite{yakhot-stokes}
theory is that the turn point in $\gamma_n$ occurs when $\tau\approx 1/\omega$.  Thus our experiments provided a direct
and unique way to extract $\tau$ as a function of pressure $p$:  Fig.~{\ref{gamma-p}}(d) displays the experimentally
extracted $\tau$ from the transition points of multiple resonators as a function of pressure. Through linear fitting,
one can obtain the expression $\tau\approx 1850/p$ [in units of nanoseconds when $p$ is in units of Torr]. The end
result of this exercise are the self-consistent fits in Fig.~{\ref{gamma-p}}(a), (b) and (c). To improve the fits, the
results emerging from Eq.~{\ref{gamma_eq}} with the appropriate material properties and $\tau$ were multiplied by
$2.8$. In general, {\it all} our data sets could be fit adequately using Eq.~{\ref{gamma_eq}} after multiplying by
$2.8\pm0.7$.

The fluidic dissipation in individual resonators shown in Fig.~{\ref{gamma-p}} suggests that there, indeed, is a
transition at $\omega\tau\approx1$, obtained by tuning $\tau$. Further support for this transition comes from extended
measurements in the frequency parameter space. Figure~{\ref{gamma-omega}}(a) shows $\gamma_n$ from different resonators
spanning a huge frequency range. Here, $\gamma_n$ is plotted against the resonator frequency $\omega_0$ at four
different pressures, i.e., four different $\tau$. This comparison between different devices with different sizes is
possible only after normalization of the dissipation by $S_{eff}/m$ \footnote{\label{foot} $S_{eff}\approx 2l(w+h)$; $m
\approx {\cal C}lwh\rho$; the value of ${\cal C}$ depends upon the structure geometry and the mode shape. A distributed
force approximation for calculating ${\cal C}$ was appropriate for our experimental conditions.}. The solid lines in
Fig.~{\ref{gamma-omega}}(a) are fits to Eq.~{\ref{gamma_eq}} using $\tau\approx 1850/p$. The points marked by arrows
correspond to $\omega\tau\approx1$.  Again, we have multiplied all the fits by $2.8$ as in Fig.~{\ref{gamma-p}}. This
multiplicative constant probably arises from adapting the theoretical expressions \cite{yakhot-stokes} for an infinite
plate oscillating {\it in-plane}  to the finite and rectangular resonators oscillating {\it out-of-plane}. Our
surface-to-volume normalization does not give the absolute dissipation, while it appears to be useful for comparing
different devices. The elucidation of finite size effects in complex geometries is the subject of our ongoing
computational research. Figure~{\ref{gamma-omega}}(b) shows all the data, $\gamma_n/ \sqrt{\omega\mu\rho_f}$, from all
devices collapsed onto a single curve, plotted against over four decades of the dimensionless parameter
$Wi=\omega\tau$. Each symbol type in Fig.~{\ref{gamma-omega}}(b) corresponds to a separate resonator; the $Wi$ is
obtained by multiplying the resonator frequency by the corresponding $\tau$ from $\tau\approx 1850/p$. The solid curve
is obtained from Eq.~{\ref{gamma_eq}}.

\begin{figure}[htb]
\includegraphics[trim=5mm 6mm 5mm 10mm,clip=true,scale=.63]{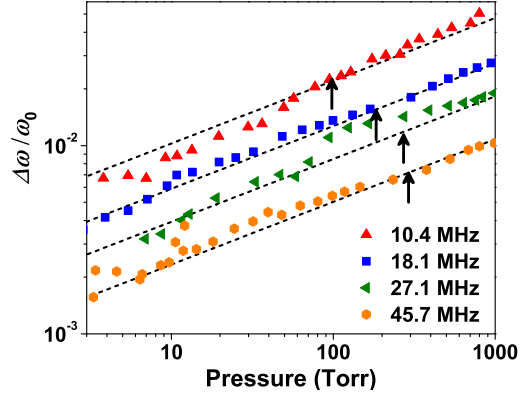}
\caption{Normalized frequency $\frac{{\Delta\omega}}{{\omega _0}}$ for beams of {\it identical} widths $w=500$ nm and
thicknesses $h=280$ nm, but varying lengths and resonance frequencies. The pressure for $Wi \approx 1$ is marked with
an arrow . Dashed lines are least-mean-squares fits to ${\Delta \omega}\propto p^{1/3}$.} \label{freq}
\end{figure}

Also apparent in Fig.~{\ref{signal}} is a small decrease in the resonance frequency as the pressure is increased. The
observed decrease is primarily due to the mass of the fluid $m_f$ that is being displaced in-phase with the resonator:
$\frac{{\Delta \omega}}{{\omega _0}} \approx \frac{{m_f}}{{2m}}$ \cite{ekinci-nems}.  Both the geometry and the
frequency of the moving surface is expected to play a role in determining $m_f$ and thus ${\Delta \omega}$. In an
effort to rule out the geometry effects, we studied the frequency shift in four beams of {\it identical} widths
($w=500$ nm) and thicknesses ($h=280$ nm) but varying lengths and resonance frequencies as a function of pressure
(Fig.~{\ref{freq}}). Here, the plotted ${\Delta \omega/ \omega_0}$  corresponds to the approximate {\it fluid mass per
unit beam length}. The arrows mark $Wi \approx 1$ for each beam and the dashed lines represent best line fits
corresponding to ${\Delta \omega}\propto p^{1/3}$. The molecular flow model \cite{bhiladvala}, which is appropriate at
low pressure, does not predict any mass loading and frequency shift. At high pressure, the  Stokes' expression for an
oscillating plate \cite{landaulifshitz} can be used to obtain an approximation for the boundary layer thickness,
$\delta = \sqrt{\frac{2\mu}{\rho_f\omega}}$. This, however, results in a pressure independent ${\Delta \omega}$, given
that typical $\delta \sim 1-5$ $\mu$m, and consequently, $\frac{{m_f}}{{m}}\approx \frac {\rho_f\delta^2}{\rho_s wh}$.
The Yakhot-Colosqui theory \cite{yakhot-stokes} underestimates the magnitude of the observed frequency shift. We expect
to understand the nature of the boundary layer in the near future by studying the scaling of experimental frequency
shifts in a wide range of geometries and frequencies, and through supporting computational analyses.

The transition observed in our experiments can be interpreted in most general terms as follows. The simple linear
relation between stress and rate-of-strain in a Newtonian fluid breaks down at high frequencies. The Boltzmannian
theory developed by Yakhot and Colosqui suggests that this result is {\it independent} of the nature of the fluid. This
transition at $Wi=\omega\tau\approx 1$ was described \cite{yakhot-stokes} as a ``viscoelastic to elastic'' transition
owing to the fact that the waves generated in the fluid by the resonator motion do {\it not} decay as
$\omega\rightarrow \infty$. There also appears to be some universality with respect to device geometry: both in
cantilevers and doubly-clamped beams, the same naive geometric normalization resulted in a consistent analysis.

There is a relentless effort to develop nanomechanical resonators operating in gaseous \cite{roukes-cantilever} and
liquid environments \cite{nems-liquid}. Our results should impact the design of next-generation nanomechanical
resonators. Figure~{\ref{gamma-omega}} suggests that fluidic dissipation saturates at high frequencies. Take, for
instance, two doubly-clamped beam resonators with identical widths and thicknesses but different lengths, i.e.,
identical $\frac {S_{eff}}{m} \approx \frac {1}{w}+\frac{1}{h}$ but different frequencies such that
$\omega_1<\omega_2$. If $\omega_1<\omega_2<1/\tau $, the ratio of the quality factors of the two resonators in fluid
becomes $\frac {Q_{2f}}{Q_{1f}} \sim \sqrt{\frac {\omega_2}{\omega_1}}$. On the other hand, if $\omega_1<1/\tau
<\omega_2 $, then $\frac {Q_{2f}}{Q_{1f}}\sim {\omega_2}\sqrt{\frac {\tau}{\omega_1}}$. Finally, for
$1/\tau<\omega_1<\omega_2$, $\frac {Q_{2f}}{Q_{1f}}\sim \frac {\omega_2}{\omega_1}$. Thus, shorter, higher-frequency
resonator will always be more resilient in a given fluid but the degree of resilience depends upon the fluid $\tau$.
Yet, for two devices with identical frequencies, the smaller one with the larger $\frac {S_{eff}}{m}$ will have the
lower $Q_{f}$. For the case where both $\frac {S_{eff}}{m}$ and device frequency increase, the nature of the scaling
determines the end result. Finally, the surface roughness, especially for very small devices, is expected to have an
important role in nanofluidics of nanomechanical resonators \cite{roughness}.

We thank M. Paul, A. Vandelay, C. Colosqui and R. Bhiladvala for helpful conversations. We acknowledge generous support
from NSF through grant Nos. CMS-324416 and BES-216274.

\bibliography{references} 
\end{document}